\documentclass[prl,twocolumn,showpacs,amsmath,amssymb]{revtex4}

\usepackage{graphicx,color}
\usepackage{dcolumn}
\usepackage{bm}
\usepackage{amsthm}
\usepackage{comment}

\newcommand{\bracket}[1]{{\langle #1 \rangle}}

\DeclareMathOperator{\trace}{Tr}
\newcommand{\Trace}[1]{{\trace[ #1 ]}}
\newcommand{\TRace}[1]{{\trace\left[ #1 \right]}}

\DeclareMathOperator{\diag}{diag}
\DeclareMathOperator{\Var}{Var}
\DeclareMathOperator{\supp}{supp}

\newcommand{\RHO}{{\hat\rho}}
\newcommand{\THeta}{{\bm{\theta}}}

\newcommand{\SIGMA}{{\hat{\bm{\sigma}}}}
\newcommand{\LAMBDA}{{\hat{\bm{\lambda}}}}

\newcommand{\qo}{{\mathcal{E}}}
\newcommand{\dualqo}{{\mathcal{E}^\dagger}}

\newcommand{\trans}{{\mathrm{T}}}

\begin{document}


\title{Optimal Measurement on Noisy Quantum Systems}
\author{Yu Watanabe$^1$}
\author{Takahiro Sagawa$^1$}
\author{Masahito Ueda$^{1,2}$}
\affiliation{$^1$Department of Physics, University of Tokyo,
7-3-1, Hongo, Bunkyo-ku, Tokyo 113-0033, Japan \\
$^2$ERATO Macroscopic Quantum Control Project, JST, 2-11-16 Yayoi, Bunkyo-ku, Tokyo 113-8656, Japan
}
\date{\today}

\begin{abstract}
We identify the optimal measurement for obtaining information about the original quantum state 
after the state to be measured has undergone partial decoherence due to noise.
We quantify the information that can be obtained by the measurement in terms of the Fisher information 
and find its value for the optimal measurement.
We apply our results to a quantum control scheme based on a spin-boson model.
\end{abstract}


\maketitle

The most serious obstacle against realizing quantum computers and networks is decoherence
that acts as a noise and causes information loss.
Decoherence occurs when a quantum system interacts with its environment, and it is unavoidable in almost all quantum systems.
Therefore, one of the central problems in quantum information science concerns 
the optimal measurement to retrieve information about the original quantum state from the decohered one and 
the maximum information that can be obtained from the measurement.

In this Letter, we identify an optimal quantum measurement that retrieves the maximum information about the expectation value 
of an observable of $\RHO$ from the partially decohered state.
Here, $\RHO$ is an unknown quantum state and modeling of the noise is assumed to be given.
The information content that we use is the Fisher information~\cite{bib:holevo:book,bib:helstrom:estimation}, 
which has been widely used in estimation theory and 
is related to the precision of the estimation.
For cases in which the unknown quantum state can be described by a single parameter, 
an optimal procedure to estimate this parameter has already been found~\cite{bib:helstrom:estimation} and 
used for phase estimation~\cite{bib:walmsley:phase-est}.
In general, a quantum state is described by multiple parameters.
The optimal estimation procedures for the multiparameter case have been discussed in several models of 
quantum systems~\cite{bib:multi-parameter-est}
and these are deeply connected with the uncertainty relations of non-commutable operators~\cite{bib:uncertainty}.
The main result of the present study is to identify the optimal measurement for a noisy quantum system
(see also \cite{bib:descrimination}).
Here, by optimal, we imply that the Fisher information obtained by the measurement is maximal and 
that the precision of the estimation from the measurement outcomes is also maximal.
While the aim of quantum error correction~\cite{bib:nielsen:error-correction} 
is to protect the unknown quantum state from interacting with the environment,
our aim is to extract maximum information from the noisy quantum system.

The crucial observation for obtaining our results is that the quantum state, observables, and quantum measurements are all described 
by a common set of generators of the Lie algebra.
This fact greatly facilitates the analysis carried out in the present study.
The Fisher information describes the precision of the parameter estimation and 
it is defined through the parameterization of quantum states.
We use a generalized Bloch vector~\cite{bib:kimura:bloch} as the parameter.
Any quantum state of a finite $N$-dimensional quantum system is expressed in terms of generators 
$\LAMBDA = \{\hat \lambda_i\}_{i=1}^{N^2-1}$ of the Lie algebra $\mathfrak{su}(N)$.
Let the generalized Bloch vector $\THeta\in\mathbb{R}^{N^2-1}$ be defined as the coefficient vector of the expansion of $\RHO$ by $\LAMBDA$:
\begin{equation}
  \RHO = \tfrac{1}{N}\hat I + \tfrac{1}{2}\THeta\cdot\LAMBDA,
\end{equation}
where $\hat I$ is the identity operator.
Since $\RHO$ is unknown, $\THeta$ is also unknown.
The generators $\LAMBDA$ satisfy 
$\hat\lambda_i^\dagger = \hat\lambda_i$, $\trace\,\hat\lambda_i = 0$, and $\Trace{\hat\lambda_i\hat\lambda_j} = 2\delta_{ij}$,
and each $\LAMBDA$ is characterized by the structure constants $f_{ijk}$ (completely antisymmetric tensor) 
and $g_{ijk}$ (completely symmetric tensor) as
$[\hat\lambda_i, \hat\lambda_j] = 2i\sum_k f_{ijk}\hat\lambda_k$,
$\{\hat \lambda_i, \hat\lambda_j\} = \frac{4}{N}\delta_{ij}\hat I + 2\sum_k g_{ijk}\hat\lambda_k$,
where $[\ ,\ ]$ and $\{\ ,\ \}$ denote the commutator and the anti-commutator, respectively.

The quantum noise in a finite-dimensional quantum system can be described as an affine map 
$\qo$~\cite{bib:qo},
$\qo(\RHO) \equiv \sum_i \hat M_i \RHO \hat M_i^\dagger$,
where $\{\hat M_i\}$ are the Kraus operators that satisfy $\sum_i \hat M_i^\dagger\hat M_i = \hat I$.
The Bloch vector $\THeta$ is also affine-mapped by $\qo$. 
By assuming that the dimension of the decohered state $\qo(\RHO)$ is the same as that of $\RHO$,
\begin{equation}
  \qo(\RHO) = \tfrac{1}{N}\hat I + \tfrac{1}{2}(A\THeta + \bm{c})\cdot\LAMBDA, \label{eq:qo_rho}
\end{equation}
where $A$ is an $(N^2-1) \times (N^2-1)$ real matrix whose $ij$-element is 
$\frac{1}{2}\Trace{\hat\lambda_i\qo(\hat\lambda_j)}$
and $\bm{c}\in\mathbb{R}^{N^2-1}$ whose $i$th element is 
$\frac{1}{N}\Trace{\hat\lambda_i\qo(\hat I)}$.
We assume that $\qo$ is injective~\cite{bib:comment:injective}; then, $A$ has an inverse, 
which physically implies that $\qo(\RHO)$ is a partially (not completely) decohered state.
The observable $\hat X$ can also be expanded by $\LAMBDA$ as
$\hat X = x_0\hat I + \bm{x}\cdot\LAMBDA$,
where $x_0 \in \mathbb{R}$ and $\bm{x}\in\mathbb{R}^{N^2-1}$.
Then, the expectation value of $\hat X$ is calculated to be 
$\bracket{\hat X} = x_0 + \bm{x}\cdot\THeta$.
Therefore, estimating $\bracket{\hat X}$ is equivalent to estimating $\bm{x}\cdot\THeta$, 
and our problem reduces to finding the measurement that maximizes the Fisher information about $\bm{x}\cdot\THeta$.

We next introduce the Fisher information.
Given $n$ $(\gg 1)$ independent and identically-distributed (i.i.d.)~quantum states $\qo(\RHO)$,
we perform the same POVM (Positive Operator Valued Measure) measurement $\bm{E}=\{\hat E_i\}_i$ on each of them.
The probability distribution of measurement outcomes is given by $p_i = \Trace{\qo(\RHO)\hat E_i}$.
In terms of $p_i$, the Fisher information about $\bracket{\hat X}$ obtained by $\bm{E}$ is defined as \cite{bib:frieden:book}
\begin{equation}
  J(\bm{x};\bm{E}) \equiv \left[\bm{x}\cdot J(\bm{E})^{-1}\bm{x}\right]^{-1}, \label{eq:def-fisher-x}
\end{equation}
where $J(\bm{E})$ is an $(N^2-1)\times(N^2-1)$ symmetric matrix called the Fisher information matrix, whose $ij$-element is defined as
$[J(\bm{E})]_{ij} \equiv \sum_k \frac{1}{p_k} \frac{\partial p_k}{\partial \theta_i}\frac{\partial p_k}{\partial \theta_j}$.
Since $J(\bm{E})$ has some zero eigenvalues, the inverse is defined on the support of $J(\bm{E})$, which we denote as $\supp[J(\bm{E})]$, 
and $J(\bm{x};\bm{E})$ is defined as zero if $\bm{x} \not\in \supp[J(\bm{E})]$.

The Fisher information characterizes the precision of the estimation.
The precision of the estimated value (estimator) $X^*$ of the unknown $\bracket{\hat X}$ can be measured by the variance of $X^*$.
If the estimator $X^*$ satisfies the unbiasedness condition, that is, 
if the expectation value of $X^*$ for all possible outcomes equals $\bracket{\hat X}$, 
the variance $\Var(X^*)$ satisfies the Cramer-Rao inequality:
$n\Var(X^*) \ge [J(\bm{x};\bm{E})]^{-1}$,
where $n$ is the number of the samples that we measure.
In general, the equality of the Cramer-Rao inequality is asymptotically satisfied for any POVM $\bm{E}$ 
by adopting the maximal-likelihood estimator as $X^*$.
Then, the estimation can be carried out most precisely with the measurement that maximizes $J(\bm{x};\bm{E})$.

The primary finding of our study is that the optimal measurement for obtaining the Fisher information about $\bracket{\hat X}$
is the projection measurement $\bm{P}_{\hat Y}$ corresponding to the spectral decomposition of an observable $\hat Y$ 
that is the solution to the operator equation
\begin{equation}
  \dualqo(\hat Y) = \hat X, \label{eq:operator-equation-xy}
\end{equation}
where $\dualqo(\hat Y)\equiv\sum_i\hat M_i^\dagger\hat Y\hat M_i$ is the adjoint map of $\qo$.
Since $\Trace{\qo(\RHO)\hat Y}=\Trace{\RHO\,\dualqo(\hat Y)}$,
the observable $\hat Y\equiv y_0\hat I + \bm{y}\cdot\LAMBDA$ is adjoint mapped as
$\dualqo(\hat Y) = (y_0 + \bm{y}\cdot\bm{c})\hat I + (A^\trans\bm{y})\cdot\LAMBDA$,
where $\trans$ denotes the transpose.
Because we assume that $A$ has an inverse,
the solution to \eqref{eq:operator-equation-xy} is obtained as
$\hat Y = \left(x_0 - ([A^\trans]^{-1}\bm{x})\cdot\bm{c}\right) + [(A^\trans)^{-1}\bm{x}]\cdot\LAMBDA$.
Although the Fisher information depends on the unknown quantum state $\RHO$~\cite{bib:comment:localized-model}, 
the observable $\hat Y$ is independent of $\RHO$.
Therefore, $\bm{P}_{\hat Y}$ is also independent of $\RHO$, 
and the optimal procedure to estimate $\bracket{\hat X}$ is simply performing $\bm{P}_{\hat Y}$ to the noisy system.
We also find that the maximum Fisher information about $\bracket{\hat X}$ is given by
\begin{equation}
  J(\bm{x};\bm{P}_{\hat Y}) = 
  (\Delta\hat Y)^{-2} \equiv 
  \left\{\Trace{\qo(\RHO)\hat Y^2} - \Trace{\qo(\RHO)\hat Y}^2\right\}^{-1}. \label{eq:maximum-info}
\end{equation}
We can also use quantum state estimation strategies~\cite{bib:hradil:state-est}
to estimate $\bracket{\hat X}$.
However, these strategies provide unnecessary pieces of information about the system
at the expense of decreasing the precision of the estimation of $\bracket{\hat X}$.
Therefore, to estimate the expectation value of a single observable $\bracket{\hat X}$,
performing $\bm{P}_{\hat Y}$ is the best strategy.

To prove these results,
we first show that the Fisher information about $\bracket{\hat X}$ obtained by the projection measurement of $\hat Y$
is expressed as \eqref{eq:maximum-info}.
Let $\bm{P}$ be a projection measurement.
Because the elements of $\bm{P} = \{\hat P_i\}_{i=1}^N$ are Hermitian operators, they are expanded in terms of $\LAMBDA$ as 
$\hat P_i = \frac{1}{N}\hat I + \bm{v}_i\cdot\LAMBDA$,
where $\bm{v}_i\in\mathbb{R}^{N^2-1}$.
For the completeness of the measurement, $\bm{v}_i$ must satisfy $\sum_{i=1}^N \bm{v}_i = \bm{0}$.
When we measure $\qo(\RHO)$ with $\bm{P}$,
the probability distribution of the outcomes is given by 
$p_i = \frac{1}{N} + \bm{v}_i\cdot(A\THeta+\bm{c})$.
Then, the Fisher information matrix $J(\bm{P})$ is calculated to be
$J(\bm{P}) = A^\trans KA$, where $K\equiv\sum_{i=1}^N p_i^{-1}\bm{v}_i\bm{v}_i^\trans$.
To calculate the Fisher information about $\bracket{\hat X}$, 
we need to find the inverse of $K$.
The support of $K$ is the space spanned by $\{\bm{v}_i\}_{i=1}^N$.
The inverse of $K$ for $\supp(K)$ is given by
$K^{-1} = (V^\trans)^{-1}QV^{-1}$,
where $V$ is an $(N^2-1)\times N$ matrix whose $i$th column vector is $\bm{v}_i$,
and $Q$ is an $N\times N$ symmetric matrix whose $ij$-element is $\delta_{ij}p_i - p_ip_j$.
Because $V$ is not a square matrix, we denote $V^{-1}$ as the generalized inverse matrix of $V$.
If we express the singular value decomposition of $V$ as $V=\sum_i s_i\bm{\zeta}_i\bm{\eta}_i^\trans$,
the generalized inverse $V^{-1}$ is defined as $V^{-1}\equiv \sum_i s_i^{-1}\bm{\eta}_i\bm{\zeta}_i^\trans$.
We therefore obtain 
\begin{align}
  J(\bm{x};\bm{P}) 
  &= [\bm{x}\cdot A^{-1}K^{-1}(A^\trans)^{-1}\bm{x}]^{-1} \notag \\
  &= [\bm{y}\cdot (V^\trans)^{-1}QV^{-1}\bm{y}]^{-1},
\end{align}
for $\bm{y}\equiv(A^\trans)^{-1}\bm{x}\in\supp(K)$, and $J(\bm{x};\bm{P}) = 0$ for $\bm{y}\not\in\supp(K)$.
The condition $\bm{y}\in\supp(K)$ is equivalent to the condition that 
$\bm{P}$ is the projection measurement $\bm{P}_{\hat Y}$
that corresponds to the spectral decomposition of an observable $\hat Y\equiv\bm{y}\cdot\LAMBDA$.
By denoting the spectral decomposition of $\hat Y$ as $\hat Y = \sum_{i=1}^N \alpha_i \hat P_i$,
it follows from the definition of $V$ and the completeness conditions of $\bm{P}$
that the $i$th eigenvalue $\alpha_i$ is equal to the $i$th element of $V^{-1}\bm{y}\in\mathbb{R}^N$.
Therefore, the Fisher information obtained from $\bm{P}_{\hat Y}$ can be calculated to be 
the inverse of the variance of $\hat Y$ on $\qo(\RHO)$:
\begin{equation}
  J(\bm{x};\bm{P}_{\hat Y}) 
  = \left[\sum_{i=1}^N\alpha_i^2 p_i - \left(\sum_{i=1}^N\alpha_i p_i\right)^2\right]^{-1} 
  = (\Delta \hat Y)^{-2}. \label{eq:classical_fisher_projection}
\end{equation}

We next show that \eqref{eq:classical_fisher_projection} gives the maximal Fisher information.
To show this, we use the quantum Fisher information~\cite{bib:helstrom:quantum-fisher}
and the quantum Cramer-Rao inequality~\cite{bib:caves:quantum-cramer-rao}.
The quantum Fisher information matrix $J^Q$ is 
independent of measurements, depends only on the measured quantum state $\qo(\RHO)$,
and gives an upper bound on the classical Fisher information matrix 
via the quantum Cramer-Rao inequality:
\begin{equation}
  J(\bm{E}) \le J^Q,\quad \text{for all }\bm{E}. \label{eq:quantum-cramer-rao-matrix}
\end{equation}
Therefore, the classical Fisher information $J(\bm{x};\bm{E})$ is bounded from above as
\begin{equation}
  J(\bm{x};\bm{E}) \le J^Q(\bm{x}), \quad \text{for all } \bm{E} \text{ and } \bm{x}, \label{eq:quantum-cramer-rao}
\end{equation}
where $J^Q(\bm{x})\equiv [\bm{x}\cdot (J^Q)^{-1}\bm{x}]^{-1}$ is the quantum Fisher information about $\bracket{\hat X}$.
Among several types of quantum Fisher information matrices that satisfy \eqref{eq:quantum-cramer-rao-matrix},
we adopt the symmetric logarithmic derivative (SLD) Fisher information matrix, 
which gives the tightest bound~\cite{bib:petz:sld-min} on \eqref{eq:quantum-cramer-rao-matrix},
whose $ij$-element is defined as
$[J^Q]_{ij} \equiv \TRace{\frac{\partial \qo(\RHO)}{\partial\theta_i} \hat L_j}$,
where $\hat L_i$ is a Hermitian operator called the SLD operator.
The SLD operator is given as the solution to 
$\frac{\partial}{\partial\theta_i}\qo(\RHO) = \frac{1}{2}\{\qo(\RHO), \hat L_i\}$.
Expanding the SLD operator as $\hat L_i = a_i\hat I + \bm{b}_i\cdot\LAMBDA$, 
from \eqref{eq:qo_rho}, we obtain 
$\bm{b}_i = \left(\frac{2}{N} I + G_{\qo(\THeta)} - \qo(\THeta)\qo(\THeta)^\trans\right)^{-1} A\bm{e}_i$
and $a_i = -\bm{b}_i\cdot\qo(\THeta)$,
where $\bm{e}_i$ is a unit vector whose $i$th element is $1$, $\qo(\THeta)\equiv A\THeta + \bm{c}$, 
and $G_{\qo(\THeta)}$ is a matrix whose $ij$-element is $\sum_k g_{ijk}\qo(\THeta)_k$.
From the definition of the SLD Fisher information matrix, 
its $ij$-element is calculated to be $(A\bm{e}_i)\cdot\bm{b}_j$.
We thus obtain
$J^Q = A^\trans\left(\frac{2}{N} I + G_{\qo(\THeta)} - \qo(\THeta)\qo(\THeta)^\trans\right)^{-1} A$.
Since we assume that $A$ has an inverse, the SLD Fisher information about $\bracket{\hat X}$ is 
\begin{align}
  J^Q(\bm{x})
  &= [\bm{x}\cdot A^{-1}(\tfrac{2}{N} I + G_{\qo(\THeta)} - \qo(\THeta)\qo(\THeta)^\trans)(A^\trans)^{-1}\bm{x}]^{-1} \notag \\
  &= \left\{\Trace{\qo(\RHO)\hat Y^2} - \Trace{\qo(\RHO)\hat Y}^2\right\}^{-1}. \label{eq:sld_fisher}
\end{align}
Then, it follows from \eqref{eq:classical_fisher_projection} 
that the projection measurement $\bm{P}_{\hat Y}$ of $\hat Y=\bm{y}\cdot\LAMBDA$
satisfies the equality of \eqref{eq:quantum-cramer-rao} 
and that $\bm{P}_{\hat Y}$ is the optimal measurement for obtaining the Fisher information about $\bracket{\hat X}$ from $\qo(\RHO)$.

Since $\Delta\hat Y$ and $\bm{P}_{\hat Y}$ are invariant under transformation $\hat Y\rightarrow y_0\hat I + \hat Y$ for any $y_0\in\mathbb{R}$,
we can choose the observable $\hat Y$ so as to satisfy $\dualqo(\hat Y)=\hat X$.
Therefore, to estimate $\bracket{\hat X}$ from the decohered state $\qo(\RHO)$,
the optimal method is to perform the projection measurement $\bm{P}_{\hat Y}$
of $\hat Y$ that satisfies the operator equation $\dualqo(\hat Y)=\hat X$.

\begin{figure}[t]
  \centering
  \includegraphics[width=246pt]{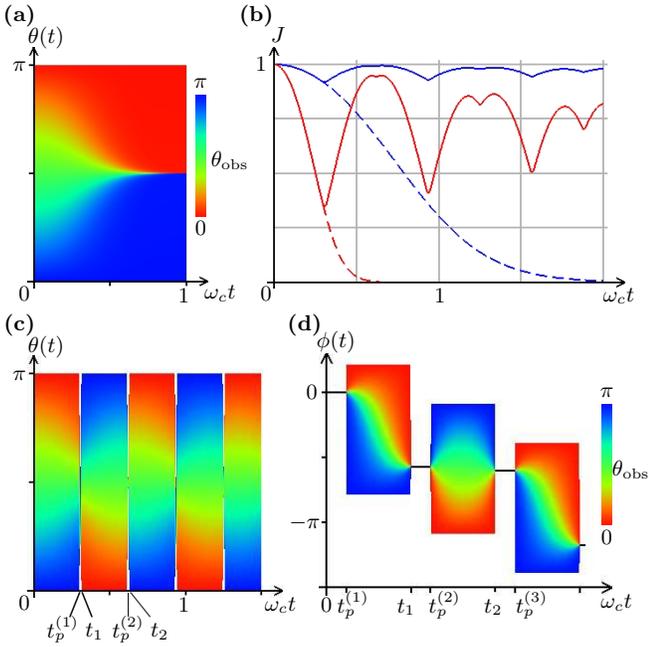}
  \caption{(Color)
    \textbf{(a)}Time evolution of $\theta(t)$ without pulse irradiation for $k_{\text{B}}T=10\hbar\omega_c$.
    \textbf{(b)}Time evolution of the maximum Fisher information $J$ about $\hat X$
    with $\theta_{\text{obs}} = 0.25\pi$ for $\RHO=\frac{1}{2}\hat I$, $\Delta t=0.3\omega_c^{-1}$, and $\tau=0.05\Delta t$.
    The red (blue) solid curve shows the high (low) temperature case with $k_{\text{B}}T = 10\hbar\omega_c$ ($k_{\text{B}}T = \hbar\omega_c$) 
    with the sequence of pulses, and the dashed curves shows the case without pulses.
    \textbf{(c)} and \textbf{(d)}Time evolutions of $\theta(t)$ and $\phi(t)$ of the optimal measurement when the sequence of pulses is applied, 
    where $k_{\text{B}}T=10\hbar\omega_c$, $\Delta t=0.3\omega_c^{-1}$, and $\tau=0.05\Delta t$.
    In (d), the time scale of pulse irradiation is magnified for clarity.}
  \label{fig}
\end{figure}

As an illustrative application of our results, let us consider a situation in which a single qubit interacts with a heat bath of bosons~\cite{bib:seth:bang-bang}.
The total Hamiltonian $\hat H_0$ is 
\begin{equation*}
  \hat H_0 = \hbar\omega_0\tfrac{\hat\sigma_z}{2} + \sum_k\hbar\omega_k\hat b_k^\dagger\hat b_k + \sum_k\hbar\hat\sigma_z(g_k\hat b_k^\dagger + g_k^*\hat b_k),
\end{equation*}
where $\hat b_k$ ($\hat b_k^\dagger$) is the bosonic annihilation (creation) operator of the heat bath.
We assume that the state of the qubit and the bath is separable at $t=0$ 
and that the initial state of the qubit is $\RHO$ and that of the bath obeys the canonical distribution.
The state of the qubit at $t$ is calculated in the interaction picture to be
\begin{equation}
  \qo(\RHO;t) = \tfrac{1}{2}(1+e^{-\Gamma_0(t)})\RHO + \tfrac{1}{2}(1-e^{-\Gamma_0(t)})\hat\sigma_z\RHO\hat\sigma_z, \label{eq:rho_t}
\end{equation}
where $\Gamma_0(t)$ increases monotonically from zero at $t=0$:
$\Gamma_0(t) \equiv 4\int_0^\infty d\omega D(\omega)\frac{1-\cos\omega t}{\omega^2}\coth\left(\frac{\beta\hbar\omega}{2}\right)$.
Here, $D(\omega)$ is the spectral density function of the bath
that we assume to take the form
$D(\omega) = \frac{1}{4}\omega\,e^{-\omega/\omega_c}$,
where $\omega_c$ is the Debye cut-off frequency.
Then, $A(t)$ and $\bm{c}(t)$ of this quantum operation \eqref{eq:rho_t} are found to be
$A(t) = \diag(e^{-\Gamma_0(t)}, e^{-\Gamma_0(t)}, 1)$ and $\bm{c}(t)=0$.
Therefore, $A(t)$ has an inverse for $t<+\infty$.
At $t=+\infty$, the right singular vector corresponding to the non-zero singular value of $A(t)$ becomes $(0, 0, 1)^\trans$; 
then, the Fisher information about all but $\hat\sigma_z$ vanishes.
If we substitute $\hat X=\sin\theta_{\text{obs}}\hat\sigma_x + \cos\theta_{\text{obs}}\hat\sigma_z$, then 
the solution to \eqref{eq:operator-equation-xy} is 
$\hat Y(t) = e^{\Gamma_0(t)}\sin\theta_{\text{obs}}\hat\sigma_x + \cos\theta_{\text{obs}}\hat\sigma_z$, 
so that the optimal measurement for $\qo(\RHO;t)$ is the projection measurement
$\hat P_{\pm}(t)=\frac{1}{2}\hat I\pm\frac{1}{2}(\sin\theta(t)\hat\sigma_x+\cos\theta(t)\hat\sigma_z)$,
where $\theta(t)$ satisfies $\tan\theta(t)=e^{\Gamma_0(t)}\tan\theta_{\text{obs}}$.
Thus, the measurement direction tilts toward the $x$-direction and eventually converges to the $x$-direction, 
as shown in Fig.~\ref{fig}(a).
Moreover, the information about $\hat X$ except for $\theta_{\text{obs}}=0$ converges to zero;
therefore, we cannot estimate any observable except for $\hat\sigma_z$ at $t=+\infty$ (see dashed curves on Fig.~\ref{fig}(b)).

In the above example, the qubit is decohered and the information about the system decreases monotonically because of the effect of the noise 
caused by the interaction with the heat bath.
It is known that the decoherence for the spin is suppressed by the spin echo technique 
by applying a sequence of pulses~\cite{bib:seth:bang-bang,bib:seth:dynamical-decoupling}.
In this case, however, $A(t)$ is not diagonal, and the measurement direction is drastically changed.
We consider the case in which the total Hamiltonian is $\hat H = \hat H_0 + \hat H_{\text{rf}}(t)$,
where $\hat H_{\text{rf}}(t)$ describes the effect of the pulse irradiations~\cite{bib:seth:bang-bang}:
$\hat H_\text{rf}(t) = \sum_k V^{(k)}(t)\{\cos[\omega_0(t - t_p^{(k)})]\hat \sigma_x + \sin[\omega_0(t - t_p^{(k)})]\hat \sigma_y\}$
with $t_p^{(k)} = k\Delta t + (k-1)\tau$ and $V^{(k)}(t) = V$ for $t_p^{(k)} \le t \le t_p^{(k)} + \tau \equiv t_k$
and $0$ otherwise.
Each pulse is applied from $t_p^{(k)}$ to $t_k$,
and the time interval to the next pulse is $\Delta t$.
Here, amplitude $V$ and duration $\tau$ are tuned to satisfy $\frac{V\tau}{\hbar}=\frac{\pi}{2}$.
Figures~\ref{fig}(c) and (d) show the change in the measurement direction 
$\bm{n}(t) = (\sin\theta(t)\cos\phi(t),\sin\theta(t)\sin\phi(t),\cos\theta(t))$ 
of the optimal measurement $P_{\pm}(t)=\tfrac{1}{2}(\hat I \pm \bm{n}(t)\cdot\SIGMA)$
for obtaining the information about $\hat X$.
The solid curves in Fig.~\ref{fig}(b) show the maximum Fisher information about an observable.
By applying the sequence of pulses, most of the lost information is recovered; 
thus, the decoherence is suppressed.

Here, we compare our optimal method with the quantum state tomography strategy~\cite{bib:tomography}.
For the example described above, the Fisher information obtained by our optimal measurement 
is three times larger than that obtained by the measurement proposed in \cite{bib:tomography}.
This is because the quantum state tomography strategy divides a given set of samples 
for use to determine three noncommutable observables, 
whereas our strategy use all of them to determine a single observable.

In conclusion, we identified an optimal method for estimating the expectation value $\bracket{\hat X}$ from a noisy quantum system.
The optimal measurement that maximizes the Fisher information is the projection measurement $\bm{P}_{\hat Y}$
corresponding to the spectral decomposition of $\hat Y$
that satisfies $\dualqo(\hat Y)=\hat X$.
We also find that the maximum Fisher information obtained by the measurement 
is given by the inverse of the variance of $\hat Y$ for the decohered state.
Although the Fisher information depends on the unknown quantum state,
the optimal measurement that maximizes the Fisher information is independent of the unknown quantum state.
Therefore, the optimal strategy for estimating $\bracket{\hat X}$ is to perform $\bm{P}_{\hat Y}$ on the noisy quantum system.
Our results are obtained under the assumptions that the quantum noise $\qo$ is injective 
and that the Hilbert space of the original state $\RHO$ and the decohered state $\qo(\RHO)$ have the same dimension.
The non-injectiveness of $\qo$ corresponds to the case in which the quantum state is completely decohered by the noise,
for example, at $t=+\infty$ in the previous example.
When quantum states are transferred by or stored on other media,
we can envisage situations in which the dimensions of the Hilbert space of $\RHO$ and $\qo(\RHO)$ are not equal.
Therefore, solving the problem in such situations is crucial for implementing quantum networks and memory.
The full investigation of this study will be reported elsewhere.

\begin{acknowledgments}
YW acknowledges M. Hayashi for the critical reading of the manuscript.
This work was supported by a Grant-in-Aid for Scientific Research (Grant No. 17071005) and by a Global COE program ``Physical Science Frontier'' of MEXT, Japan.
YW and TS acknowledge support from JSPS (Grant No. 216681 and No. 208038, respectively).
\end{acknowledgments}


\begin{thebibliography}{99}
\bibitem{bib:comment:injective}
If $\qo$ is not injective, the support of the Fisher information $J(\bm{E})$
is restricted to $\supp(J(\bm{E}))\subseteq \mathcal{L}_R(A)$ for all $\bm{E}$,
where $\mathcal{L}_R(A)$ is the space spanned by the right singular vectors corresponding to the non-zero singular values of $A$.
Then, the Fisher information about $\bracket{X}$ for all $\bm{x}\not\in \mathcal{L}_R(A)$ is zero for all $\bm{E}$;
therefore the variance of the estimator $X^*$ of $\bracket{\hat X}$ does not converge to zero.
\end{thebibliography}

\end{document}